%% file: ms.tex
\documentclass{article}

\usepackage{lipsum}

\newcommand\blfootnote[1]{%
  \begingroup
  \renewcommand\thefootnote{}\footnote{#1}%
  \addtocounter{footnote}{-1}%
  \endgroup
}
 
\usepackage{amsmath,amsfonts,amssymb}
\usepackage{graphicx}
\usepackage[colorlinks=true, allcolors=blue]{hyperref}

\usepackage{todonotes}
\usepackage{multirow}
\usepackage{multicol}
\usepackage{caption}
\usepackage{subcaption}
\usepackage{authblk}
\usepackage{geometry}

\newgeometry{vmargin={20mm, 20mm}, hmargin={30mm,30mm}}   

\usepackage{soul}
\usepackage{booktabs}
\usepackage{pdfpages}
\usepackage{rotating}
\usepackage{makecell}

\title{WeakSTIL: Weak whole-slide image level stromal tumor infiltrating lymphocyte scores are all you need}

\author[a, b]{Yoni Schirris*}
\author[a, c]{Mendel Engelaer*}
\author[a, b]{Andreas Panteli}
\author[a]{Hugo Mark Horlings}
\author[b, e]{Efstratios Gavves}
\author[a, b, d]{Jonas Teuwen}
\affil[a]{Netherlands Cancer Institute}
\affil[b]{University of Amsterdam}
\affil[c]{VU University Amsterdam}
\affil[d]{Radboud University Medical Center}
\affil[e]{Ellogon AI B.V.}

\begin{document} 
\maketitle
\blfootnote{*Equal contribution. Send correspondence to Jonas Teuwen. Address: Plesmanlaan 161, 1066 CX Amsterdam, the Netherlands; E-mail: j.teuwen@nki.nl}
\begin{abstract}
We present WeakSTIL, an interpretable two-stage weak label deep learning pipeline for scoring the percentage of stromal tumor infiltrating lymphocytes (sTIL\%) in H\&E-stained whole-slide images (WSIs) of breast cancer tissue. The sTIL\% score is a prognostic and predictive biomarker for many solid tumor types. However, due to the high labeling efforts and high intra- and interobserver variability within and between expert annotators, this biomarker is currently not used in routine clinical decision making. WeakSTIL compresses tiles of a WSI using a feature extractor pre-trained with self-supervised learning on unlabeled histopathology data and learns to predict precise sTIL\% scores for each tile in the tumor bed by using a multiple instance learning regressor that only requires a weak WSI-level label. By requiring only a weak label, we overcome the large annotation efforts required to train currently existing TIL detection methods. We show that WeakSTIL is at least as good as other TIL detection methods when predicting the WSI-level sTIL\% score, reaching a coefficient of determination of $0.45\pm0.15$ when compared to scores generated by an expert pathologist, and an AUC of $0.89\pm0.05$ when treating it as the clinically interesting sTIL-high vs sTIL-low classification task. Additionally, we show that the intermediate tile-level predictions of WeakSTIL are highly interpretable, which suggests that WeakSTIL pays attention to latent features related to the number of TILs and the tissue type. In the future, WeakSTIL may be used to provide consistent and interpretable sTIL\% predictions to stratify breast cancer patients into targeted therapy arms. 
\end{abstract}

\section{Introduction}
\label{sec:intro}

For many types of solid tumors, the presence and distribution of tumor infiltrating lymphocytes (TILs) as seen in Hematoxylin and Eosin (H\&E) stained Whole Slide Images (WSIs) of resected tumor tissue have a prognostic \cite{sanchez2020pdtilprognostic, maibach2020tumortilprognostic, gao2020prognostictilprognostic, idos2020prognostictilprognostic, loi2013prognostictilprognosticpredictive} and predictive \cite{loi2013prognostictilprognosticpredictive, paijens2021tumortilpredictive, lee2018prognostictilpredictive, o2018reproducibilitytilpredictive, stenzel2020prognostictilpredictive, lianyuan2018predictivetilpredictive} value. Despite guidelines by the international TILs working group \cite{hendry2017assessingtilsworkinggrouppart1, hendry2017assessingtilsworkinggrouppart2}, the annotation of TILs still requires trained pathologists, remains time-consuming, and the issue of intra- and inter-observer variability is not entirely overcome \cite{salgado2018beginning, gonzalez2020path}. These issues hamper the routine application of TILs as a biomarker for clinical decision-making. Computational methods promise a way to overcome these issues.

Current computational methods that compute TILs-based scores either annotate all TILs in the tissue using deep learning based \cite{panteli2021sparse, lu2020deeptildetection, sun2021computational, thagaard2021automatedstiltnbc} or classical computational methods \cite{chou2021optimization}, or classify each patch based on whether or not it contains TILs \cite{saltz2018spatial}. Although these methods allow the analysis of the spatial distribution of TILs, these methods require many costly annotations of TILs to be trained. Additionally, these detailed distributions may not be necessary, since one of the promising biomarkers in, e.g., breast cancer, is an aggregated percentage of TILs in the tumor stroma (stromal TILs score, or sTIL\% score) \cite{hammerl2018breaststilpercent, savas2016clinicalbreaststilpercent}. Concluding, an automated algorithm can be clinically valuable when it only produces a single WSI-level score, for which the training labels can be more easily attained than laborious manual cell-level annotations.
    
Lately, weak label learning for H\&E WSIs has shown promising results using self-supervised learning (SSL) and multiple instance learning (MIL) for tumor detection in the CAMELYON16 dataset and for genomic feature classification in The Cancer Genome Atlas (TCGA) dataset \cite{schirris2021deepsmile, dehaene2020self}. In this work, we investigate the effectiveness of SSL and MIL to predict the WSI-level sTIL\% score directly from an H\&E WSI of breast cancer tissue. We expect this method to provide consistent and explainable sTIL\% scores when trained with only weak labels that are less time-consuming to generate than exact TIL annotations.

Our main contributions can be summarized as follows:
\begin{enumerate}
    \item We propose WeakSTIL, a weak label deep learning pipeline using SSL and TILMIL. The SSL-pretrained feature extractor is used to compress the WSI, and TILMIL is a MIL regressor adjusted for the task of sTIL\% scoring in H\&E WSIs. TILMIL is trained using only weak WSI-level sTIL\% labels. 
    \item We show that an in-domain pretrained feature extractor with SSL is essential to the success of this method, and that using features extracted using an ImageNet-pretrained feature extractor leads to random performance.
    \item We show that WeakSTIL is at least as good as a high-resolution TIL detection pipeline while requiring a much lower annotation effort.
\end{enumerate}

\section{Materials and methods}
\label{sec:mm}

\paragraph{Dataset}
We use 286 Formalin-Fixed Paraffin Embedded (FFPE) digitized H\&E WSIs of breast cancer tissue from TCGA \cite{koboldt2012comprehensivetcgabc} with sTIL\% labels. The WSI-level sTIL\% labels were manually scored by an expert pathologist, following the scoring guidelines as proposed by the International TILs Working Group \cite{hendry2017assessingtilsworkinggrouppart1, hendry2017assessingtilsworkinggrouppart2}. We binarize the score of patient $i$ as TILs-low ($\text{sTIL\%}_i \leq 0.2$) or TILs-high ($\text{sTIL\%}_i > 0.2)$ for the classification task. We perform 5-fold cross-validation using a 60/20/20 train/validation/test split, where also the test set is rotated. All splits are performed on a patient level and are stratified on PAM50 breast cancer subtypes \cite{parker2009supervisedpam50}. We split the WSIs in tiles of $512\times512$px at a resolution of $0.5$ microns per pixel (mpp). We discard background tiles using the improved foreground extraction for histopathological whole slide imaging algorithm \cite{riasatian2020comparative, bug2015foregroundimprovedfesi} and only use tiles from manually annotated tumor bed regions, which include both the tumor area and tumor stromal area.

\paragraph{Baseline} As a baseline we use the results obtained from a TIL detector as described in earlier work \cite{panteli2021sparse}, which we refer to as \textit{DetecTIL} (from \textit{exhaustive Detection of TILs}). To obtain the tumor bed TIL\% estimation ($\text{tbTIL\%}$) from the detected TILs, we compute the fraction of the tumor bed area that is occupied by TILs using the number of TILs ($\text{\#TILs}$), the average area occupied by a TIL ($A_{\text{TIL}}=\pi r^2 \approx \pi 4^2 \approx 50.2 \mu m^2$, assuming an average radius of $4 \mu m$ per lymphocyte \cite{hendry2017assessingtilsworkinggrouppart1}), and the total area of the tumor bed ($A_\text{tb}=\text{\#tiles}_{\text{tb}} \times A_{\text{tile}} = \text{\#tiles}_{\text{tb}} \times (\text{w}_\text{tile}\times \text{mpp}) \times (\text{h}_\text{tile} \times mpp) = \text{\#tiles}_{\text{tb}} \times (512\times0.5)^2)$ as follows:
    \begin{equation}
        \text{tbTIL\%} = \frac{\text{\#TILs} \times A_{\text{TIL}}}{A_{\text{bed}}}
    \end{equation}
Note that exhaustive tumor bed tumor-stroma segmentation annotations are necessary for the best performance for this model, yet these annotations are not available to the authors at the time of writing. Therefore, the $\text{tbTIL\%}$ is likely a biased estimate of the $\text{sTIL\%}$ label.

\paragraph{WeakSTIL}
 We use a two-stage learning method similar to Schirris et al. \cite{schirris2021deepsmile} and Dehaene et al. \cite{dehaene2020self}. In the first stage, we compress the WSI by extracting features from all tiles using a pre-trained feature extractor. In the second stage, we perform Multiple Instance Learning (MIL) on the extracted features to predict the sTIL\% score from the compressed WSI.
   
For the feature extraction, we compare the baseline Resnet18 pre-trained on ImageNet (\textit{IN-RN18}), to a Resnet18 pre-trained with SimCLR \cite{chen2020simple} on a variety of histopathology datasets \cite{ciga2020self-ssl-for-digital-pathology} (\textit{HistoSSL-RN18}).

For the WSI-level sTIL\% score predictions, we propose \textit{TILMIL} (from \textit{TIL score prediction using Multiple Instance Learning}). In TILMIL, we use an adjusted multiple instance learning assumption. The traditional multiple instance learning setting is stated as a classification problem where the label of the bag (in our case: WSI) of instances (in our case: tiles) is positive if any one of the instances is labeled as positive. First, we state it as a regression task instead of a classification task. Second, we assume that the predictive signal of the WSI-level label is found equally in each tumor bed tile. 
Therefore, we predict a continuous sTIL\% score for each $n^{\text{th}}$ tile ($\text{sTIL\%}_n)$ from its $H$-dimensional latent feature representation, $\boldsymbol{h}_n$, using a linear neural network layer with weights $\boldsymbol{w} \in \mathbb{R}^{H}$ and bias $b \in \mathbb{R}$ (see appendix, section \ref{sec:apx-hparam}, for the evaluation of different classifier layers) after which the WSI-level sTIL\% score is the average of these tumor bed tile-level scores:

\begin{equation}
    \text{sTIL\%} = \frac{1}{N} \sum_k \text{sTIL\%}_n = \frac{1}{N} \sum_n \text{sigm} \left( \boldsymbol{w}^\intercal \boldsymbol{h}_n + b \right)
\end{equation}

We train TILMIL on IN-RN18 and HistoSSL-RN18 features with the Adam optimizer, a learning rate of $1\times10^{-2}$ and $5\times10^{-3}$, and L2 norm of $5\times10^{-4}$ and $1\times10^{-4}$, respectively. Both methods are trained for 50 epochs with a batch size of 1 (i.e. a single WSI with a varying number of tiles), and we evaluate the performance on the validation set every epoch. These hyperparameters are chosen after a hyperparameter grid search as presented in the appendix, section~\ref{sec:apx-hparam}. The loss is computed as the mean squared error of the model-estimated sTIL\% score and the pathologist-derived sTIL\% score. In our case, $H=512$, extracted using the pre-trained Resnet18 by Ciga et al. \cite{ciga2020self-ssl-for-digital-pathology}. 
    
\paragraph{Evaluation} All experiments are run with the same random seed initialization. Although we train the model with a regression task, we select the model with the highest area under the receiver operating curve (AUC) evaluated on the binarized labels of the validation set and use this model to infer the prediction for the test set. We report the AUC computed between the predicted scores and binary labels, and the Pearson's $r$ and $R^2$ between the predicted scores and the pathologist's scores, computed in python v3.9.6 with sklearn v0.24.2 and scipy v1.7.1. Note that the $R^2$ is computed without bias, $R^2=1$ means that all variability in the pathologist's score is explained by the predicted value, $R^2=0$ means that the predictive value of the predicted score is as good as the mean of the pathologist's scores, and that $R^2<0$ indicates that the predictive value of the predictions is worse than a model that would predict the mean.
\section{Results}
\label{sec:results}

Table~\ref{tab:results} displays the results of our main regression experiments, evaluated both as a regression task and as a classification task. We display the scatter plots of the predicted scores compared to the real scores in Figure~\ref{fig:results-scatter} for the HistoSSL-RN18 TILMIL, and present the ROC curves for IN-RN18 and HistoSSL-RN18 in Figure~\ref{fig:results-roc}. 

\begin{table}[!t]
    \caption{Main results of WeakSTIL, compared to DetecTIL on the tumor bed subset.} 
    \label{tab:results}
    \begin{center}       
        \begin{tabular}{l r r}
            \toprule
            \textbf{Model} & \multicolumn{1}{c}{\textbf{$R^2$}} & \multicolumn{1}{c}{\textbf{$AUC$}} \\
            \midrule
            DetecTIL \cite{panteli2021sparse} & $-0.45 \pm 0.12$ & $0.75 \pm 0.05$\\
            IN-RN18 + TILMIL & $-1.56\pm0.23$ & $0.63 \pm 0.11$ \\
            HistoSSL-RN18 + TILMIL (WeakSTIL) & $\boldsymbol{0.45\pm0.15}$ & $\boldsymbol{0.89\pm 0.05}$   \\
            \bottomrule
        \end{tabular}
    \end{center}
\end{table}

\paragraph{DetecTIL} First, we find that our TIL detection baseline reaches a performance of $0.75 \pm 0.05$ AUC for the sTIL-high versus sTIL-low classification. On the regression task, it reaches an $R^2$ of $-0.45\pm0.12$, indicating a predictive performance worse than a model that outputs the mean. These results indicate that there is an ordering in the predicted scores which correlate with the binary labels, but that the absolute predicted scores are not indicative of the real scores.

\paragraph{IN-RN18 + TILMIL} Secondly, in the weak label learning setup, it is seen that the performance of TILMIL on the ImageNet-pretrained feature extractor performs worse than DetecTIL. As seen in Table~\ref{tab:results}, IN-RN18+TILMIL reaches an AUC of $0.63\pm0.11$ and an $R^2$ of $-1.56\pm0.23$, indicating a random performance. Figure~\ref{fig:results-roc} shows the near-random ROC curves of the model. 

\paragraph{HistoSSL-RN18 + TILMIL} When using the HistoSSL pre-trained feature extractor, however, WeakSTIL reaches an AUC of $0.89\pm0.05$ on the sTIL-high versus sTIL-low classification task. On the regression task, WeakSTIL reaches an $R^2$ of $0.45\pm0.15$, which shows that the absolute scores predicted are near the real scores. This relationship between the predicted and real sTIL\% scores is visualized in the scatter plots in Figure~\ref{fig:results-scatter}, reaching a Pearson's $r$ of $0.82\pm0.05$.

\paragraph{Interpretability} Lastly, WeakSTIL produces interpretable results by providing realistic tile-level scores, as shown in Figure \ref{fig:results-visualization}. These qualitative visualizations suggest that WeakSTIL pays attention to latent features related to the number of TILs and the tissue type (stroma or tumor).

\input{scatter_plot}

\input{roc_figure}
\input{heatmap_figure}

\section{Discussion and conclusion}
\label{sec:conclusion}

For WeakSTIL, IN-RN18 features do not encode sufficient information for TILMIL to learn to score stromal TILs. When using HistoSSL-RN18 features, though, TILMIL learns to use those features to predict the sTIL\% score, confirming earlier evaluations of the effectiveness of SSL in the histopathology domain \cite{schirris2021deepsmile, dehaene2020self, ciga2020self-ssl-for-digital-pathology}. Using the SSL features, WeakSTIL outperforms DetecTIL by $0.14$ AUC on the classification task while showing a linear relationship between the predicted and true scores with an $R^2>0$.

The seemingly flawed performance of DetecTIL can be attributed to two main points. First, the ground truth sTIL\% scores are rounded estimates by pathologists, known to vary between pathologists. The actual space occupied by TILs in tumor bed stroma may thus differ from the sTIL\% score provided, and DetecTIL might perform better than the pathologist. Secondly, the authors lack tumor-stroma segmentations, which means that we compute the tumor bed TIL\% score instead of the tumor bed stromal TIL\% score, which biases the predicted score. 

Even though a perfect comparison is not possible due to different training and validation datasets, a preliminary comparison of WeakSTIL to a TIL scoring pipeline by Thagaard et al. \cite{thagaard2021automatedstiltnbc} ($r=0.79$), which is much more extensive than the baseline pipeline used in this study, indicates that WeakSTIL reaches a similar performance ($r=0.82\pm0.05$) when comparing to a method that uses precise tumor-stroma segmentation methods.

Although these preliminary results are promising, we also note three limitations of WeakSTIL. First, TILMIL computes the mean over all tumor bed tile scores, which dilutes the sTIL\% score. This leads to a failure mode especially on WSIs with a high sTIL\% score but with little stroma (see appendix, section~\ref{sec:apx-heatmap-failure-wsi}). A possible way to overcome this is by using an additional network that outputs a binarized attention weight that is essentially a tumor-stroma classifier. This intermediate output can then be used to filter out the non-stroma tiles when computing the mean. Second, since we do not use supervisory tumor-stroma segmentations or tile-level sTIL\% scores, the model may need more WSIs to learn how to interpret uncommon tiles. Currently, this leads to a tile-level failure mode where edge tiles with few TILs receive a relatively high score (see appendix, section~\ref{sec:apx-heatmap-failure-tile}). Lastly, WeakSTIL only looks at a tile-level context. Since this does not allow the model to recognize whether stromal area is located inside or outside the tumor bed, tumor bed annotations are still required.
 
In conclusion, WeakSTIL utilizes self-supervised learning and a simple MIL regressor to produce interpretable tile-level sTIL\% scores which perform at least as well as exhaustive TIL detection models while requiring fewer annotations. Weak label learning is a promising avenue for WSI-level sTIL\% predictions, given that our results suggest that weak sTIL\% labels are all you need to train a neural network to predict this prognostic and predictive biomarker directly from H\&E WSIs.

\section*{Acknowledgments}
The collaboration project is co-funded by the PPP Allowance made available by Health~Holland\footnote{\url{https://www.health-holland.com}}, Top Sector Life Sciences \& Health, to stimulate public-private partnerships.

We would like to thank Roberto Salgado from GZA-ZNA Hospitals, Antwerp, for providing us with the sTIL\% labels of the breast cancer samples of the TCGA used as a target in this study. Additionally, we would like to thank Jakob Kather and Jeremias Krause from RWTH Aachen for providing us with the
ir tumor bed annotations of the WSIs of the FFPE WSIs of the breast cancer samples of the TCGA.

\bibliography{ms} 
\bibliographystyle{spiebib} 

\clearpage

\appendix    

\section{Hyperparameter search}
\label{sec:apx-hparam}

Table \ref{tab:appendix-table-lr-reg-ssl-linear}, \ref{tab:appendix-table-lr-reg-ssl-non-linear}, \ref{tab:appendix-table-lr-reg-ssl-double-head}, and \ref{tab:appendix-table-lr-reg-imagenet-linear} show the mean and standard deviation of the highest AUC scores achieved during each fold for varying learning rates and regularization (L2 norm) values. Table~\ref{tab:appendix-table-lr-reg-ssl-linear} shows the results for the HistoSSL-RN18 extractor with the linear TILMIL model. Table~\ref{tab:appendix-table-lr-reg-ssl-non-linear} shows the results for the HistoSSL-RN18 extractor using two linear layers for the tile-level classification head. Table~\ref{tab:appendix-table-lr-reg-ssl-double-head} shows the results for the HistoSSL-RN18 extractor with a double non-linear classification head. Table~\ref{tab:appendix-table-lr-reg-imagenet-linear} shows the results for the ImageNet-RN18 extractor with the linear TILMIL model.

Since we see few differences in maximum performance for the varying classification heads on the HistoSSL-RN18 extractor, we continue with the simplest model for the final evaluations on the test set; the single linear classification head. 

\input{hyperparameter_tables}

\section{Inspection of WeakSTIL predictions}

\subsection{Tile- and WSI-level success} \label{sec:apx-heatmap-success}

Similarly to Figure~\ref{fig:results-visualization}, the cases in Figure~\ref{fig:apx-viz-success-1} and Figure~\ref{fig:apx-viz-success-2} receive a predicted score close to the pathologist's score with sensible tile-level predictions without clearly noticeable failure modes.

\input{fig-heatmap-appendix-success-1}
\input{fig-heatmap-appendix-success-2}

\subsection{WSI-level failure: Dilution of sTIL\% scores} \label{sec:apx-heatmap-failure-wsi}

Figure~\ref{fig:apx-viz-fail-high} displays an example of a failure mode of WeakSTIL. For high sTIL\% cases with a high proportion of tumor area in the tumor bed, the final sTIL\% is diluted by low sTIL\% scores in the tumor areas. Although the stromal areas with many TILs are correctly receiving high scores, the final score is too low. 

\input{fig-heatmap-appendix-fail-high-stil}

\subsection{Tile-level failure: Uncommon tiles} \label{sec:apx-heatmap-failure-tile}

Figure~\ref{fig:apx-viz-fail-edge} displays an example of a failure mode of WeakSTIL. Since most cases have an annotated tumor bed without edge tiles, these tiles are uncommon in the training set. These edge tiles are often given high scores, while it is not necessarily stromal area, and does not have any tumor infiltrate. Similar to Figure~\ref{fig:apx-viz-fail-high}, we see a relatively large tumor area, diluting the final WSI-level score.

\input{fig-heatmap-appendix-fail-edge}
\end{document}

%% file: scatter_plot.tex
\begin{figure}[!t]
    \centering
    \setlength{\tabcolsep}{0pt}
    \begin{tabular}{p{1em}ccccc}
        \rotatebox[origin=c]{90}{sTILs true} &
        \begin{subfigure}{0.18\textwidth}
            \centering
            \includegraphics[width=\textwidth]{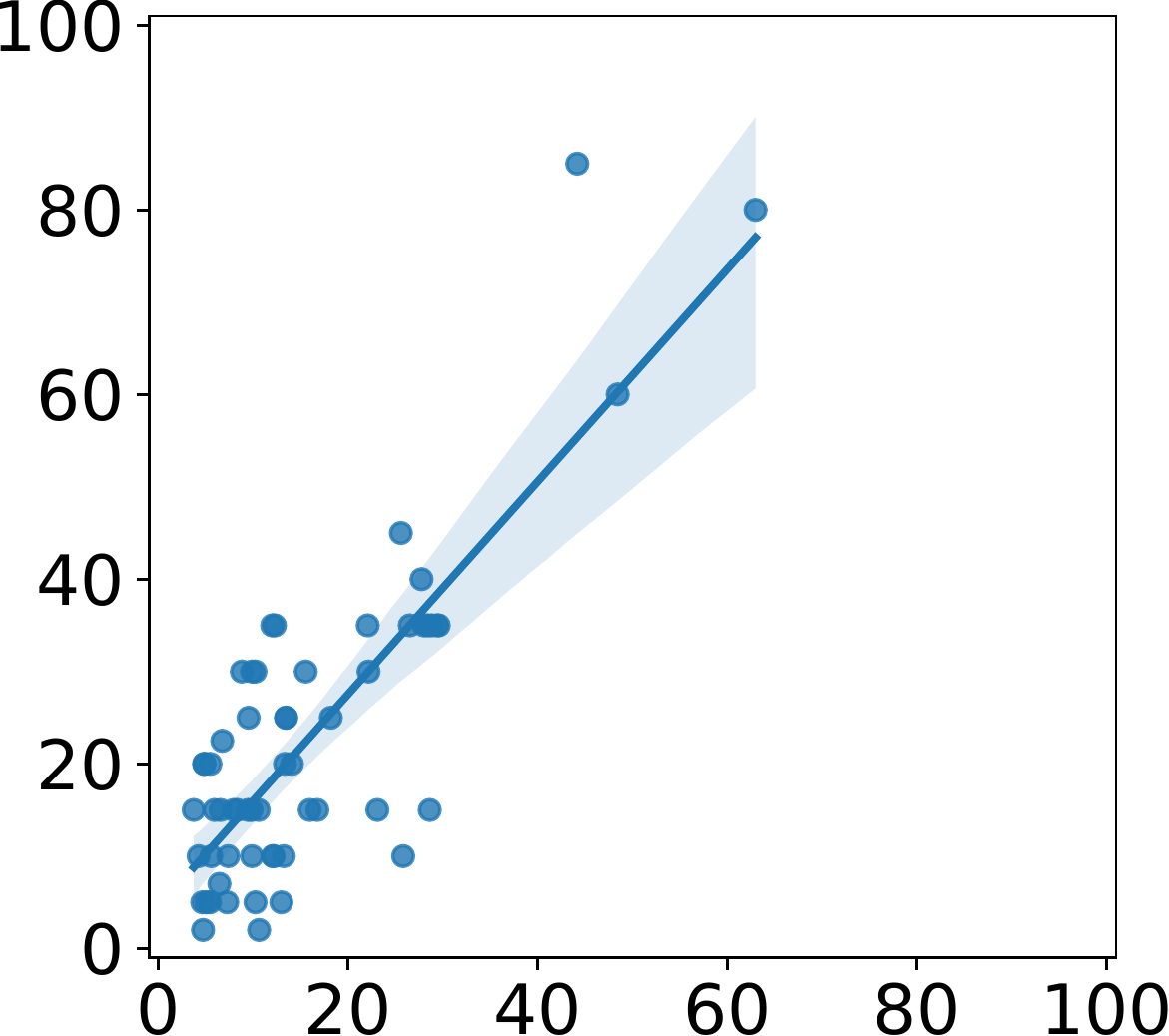}
        \end{subfigure} \vspace{0.1ex}

        &
        \begin{subfigure}{0.18\textwidth}
            \centering
            \includegraphics[width=\textwidth]{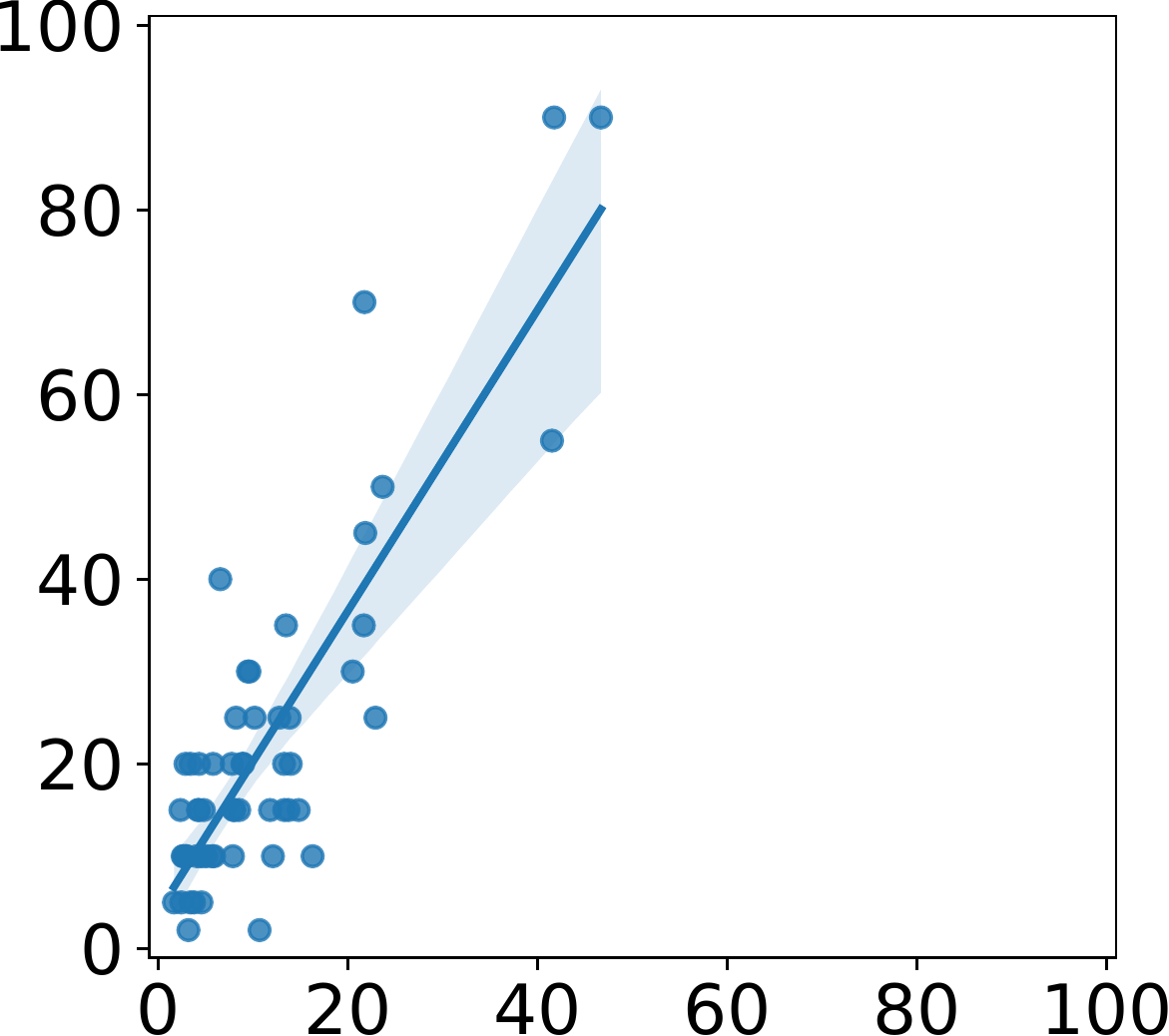}
        \end{subfigure} \vspace{0.1ex}

        &
        \begin{subfigure}{0.18\textwidth}
            \centering
            \includegraphics[width=\textwidth]{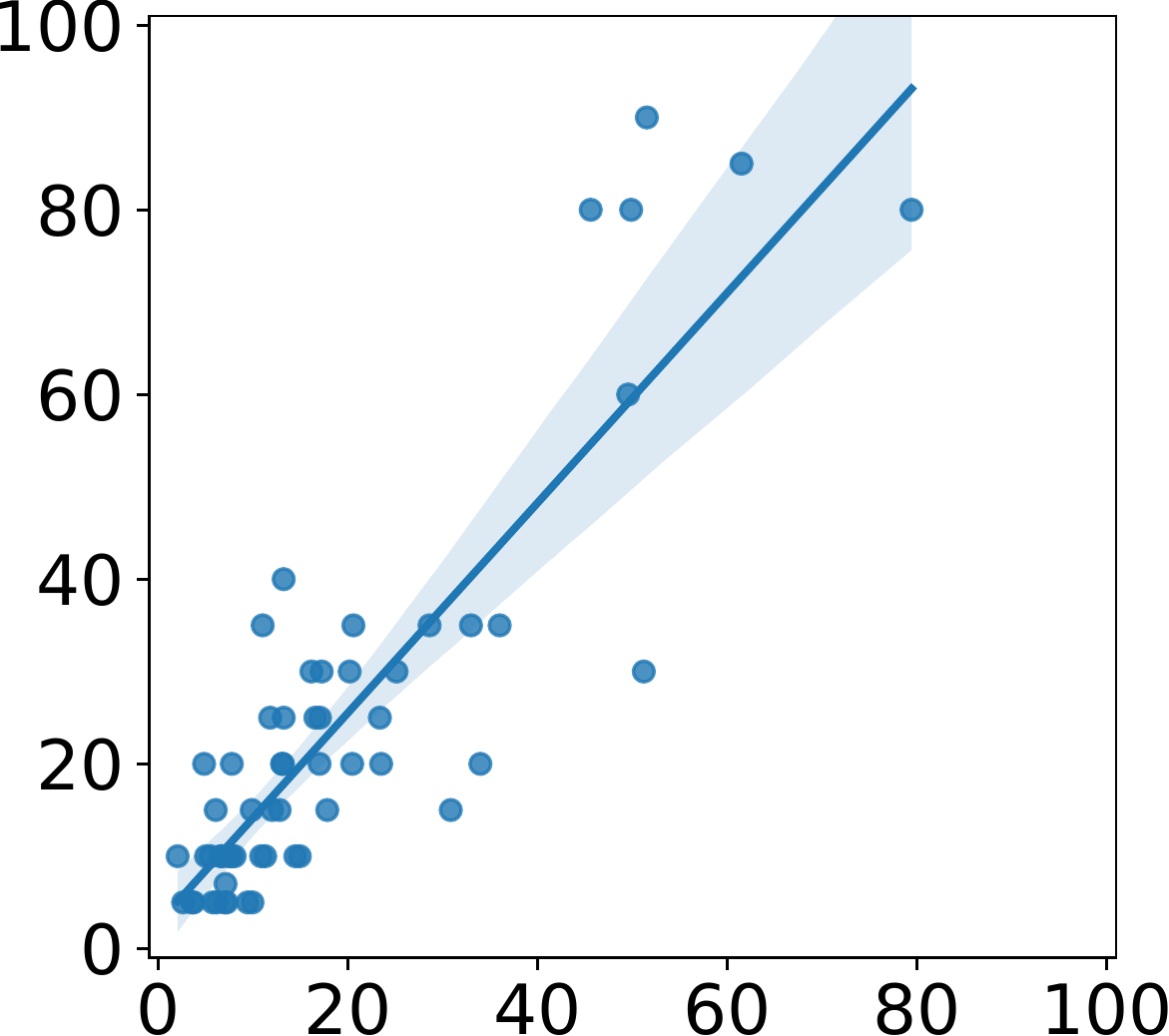}
            
        \end{subfigure} \vspace{0.1ex}
        &
        \begin{subfigure}{0.18\textwidth}
            \centering
            \includegraphics[width=\textwidth]{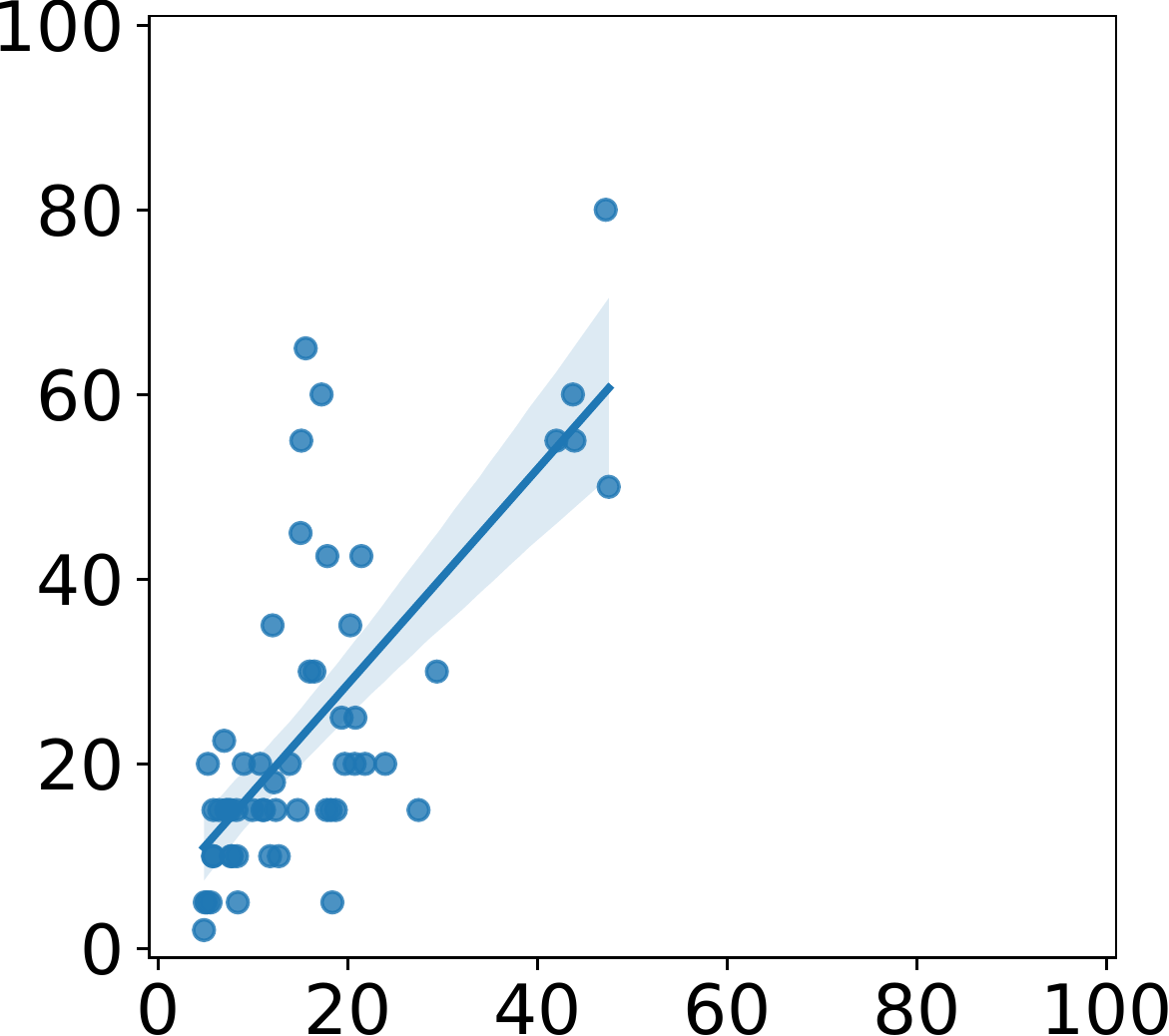}
            
        \end{subfigure} \vspace{0.1ex}
        &
        \begin{subfigure}{0.18\textwidth}
            \centering
            \includegraphics[width=\textwidth]{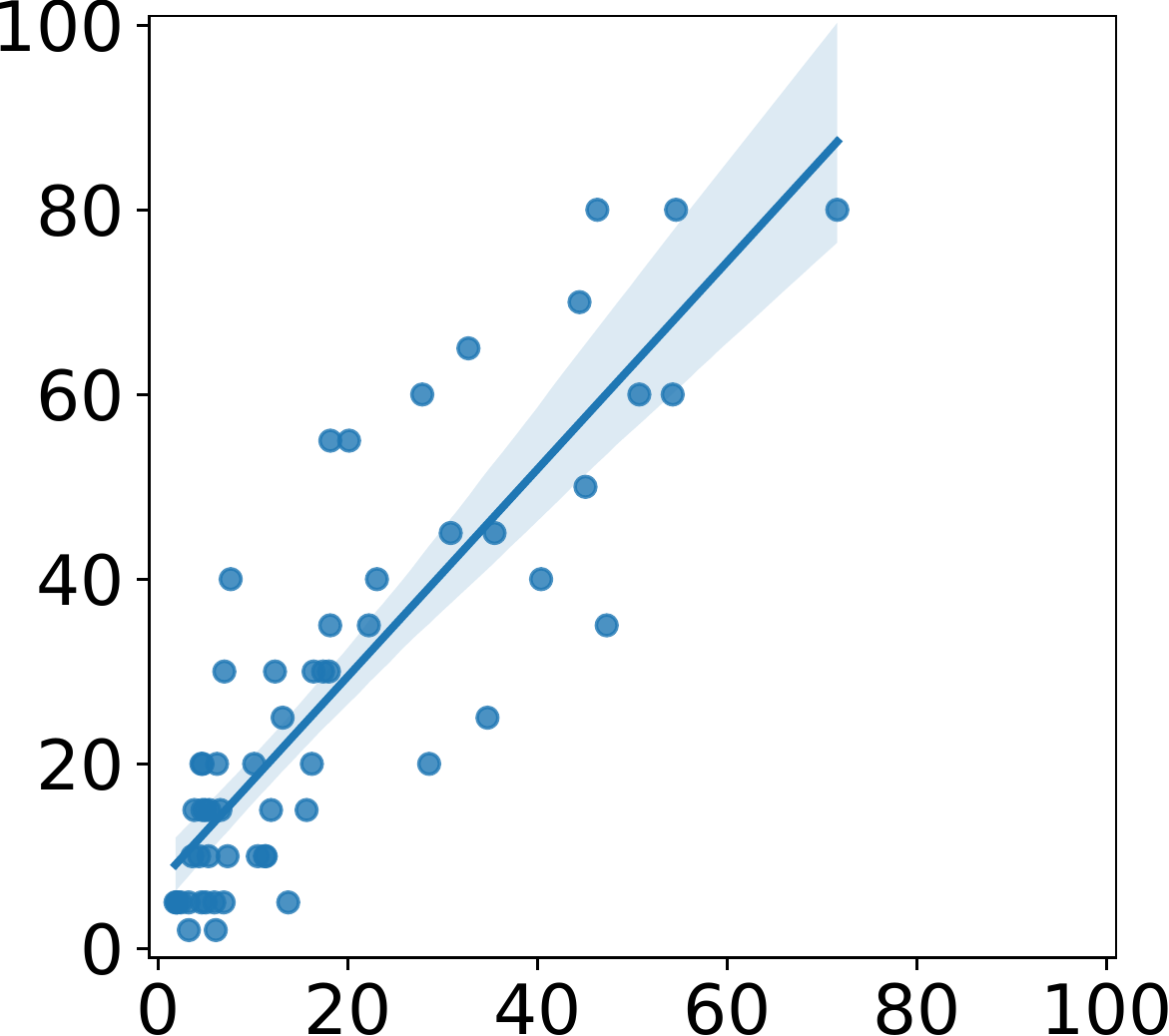}
        \end{subfigure} \vspace{0.1ex}

        \\
        & \subfloat[$r = 0.81^*$]{sTILs Predicted} & \subfloat[$r = 0.85^*$]{sTILs Predicted} & \subfloat[$r = 0.87^*$]{sTILs Predicted} & \subfloat[$r = 0.72^*$]{sTILs Predicted} & \subfloat[$r = 0.86^*$]{sTILs Predicted}
    \end{tabular}
    \caption{\label{fig:results-scatter} Scatter plots comparing the predicted WSI-level sTIL\% score to the pathologist-annotated sTIL\% score for the HistoSSL-RN18 TILMIL model for each fold. The caption states the correlation between the predicted sTIL\% score and the actual sTIL\% score. (a) Fold 1: $R^2=0.48$, (b) Fold 2: $R^2=0.28$, (c) Fold 3: $R^2=0.68$, (d) Fold 4: $R^2=0.30$, (e) Fold 5: $R^2=0.54$. $^{*}: p<0.001$}
\end{figure}

%% file: roc_figure.tex
\begin{figure}[t!]
    \centering
    \begin{tabular}{lc lc}
        & ROC for ImageNet-RN18 TILMIL
        & & ROC for HistoSSL-RN18 TILMIL\\
        \rotatebox[origin=c]{90}{True positive rate} &
        \begin{subfigure}{0.4\textwidth}
            \centering
            \includegraphics[width=.8\textwidth]{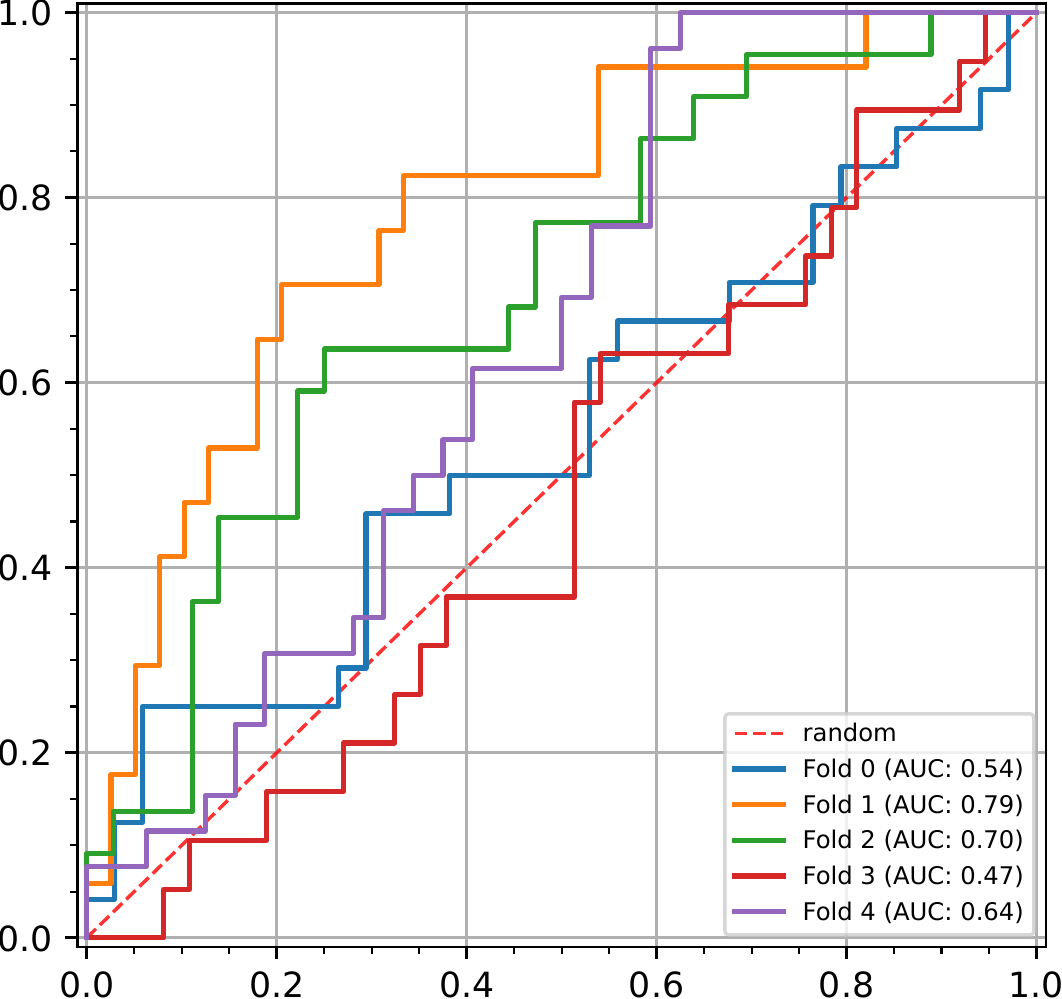}
        \end{subfigure} \vspace{0.3ex} &
         &
        \begin{subfigure}{0.4\textwidth}
            \centering
            \includegraphics[width=.8\textwidth]{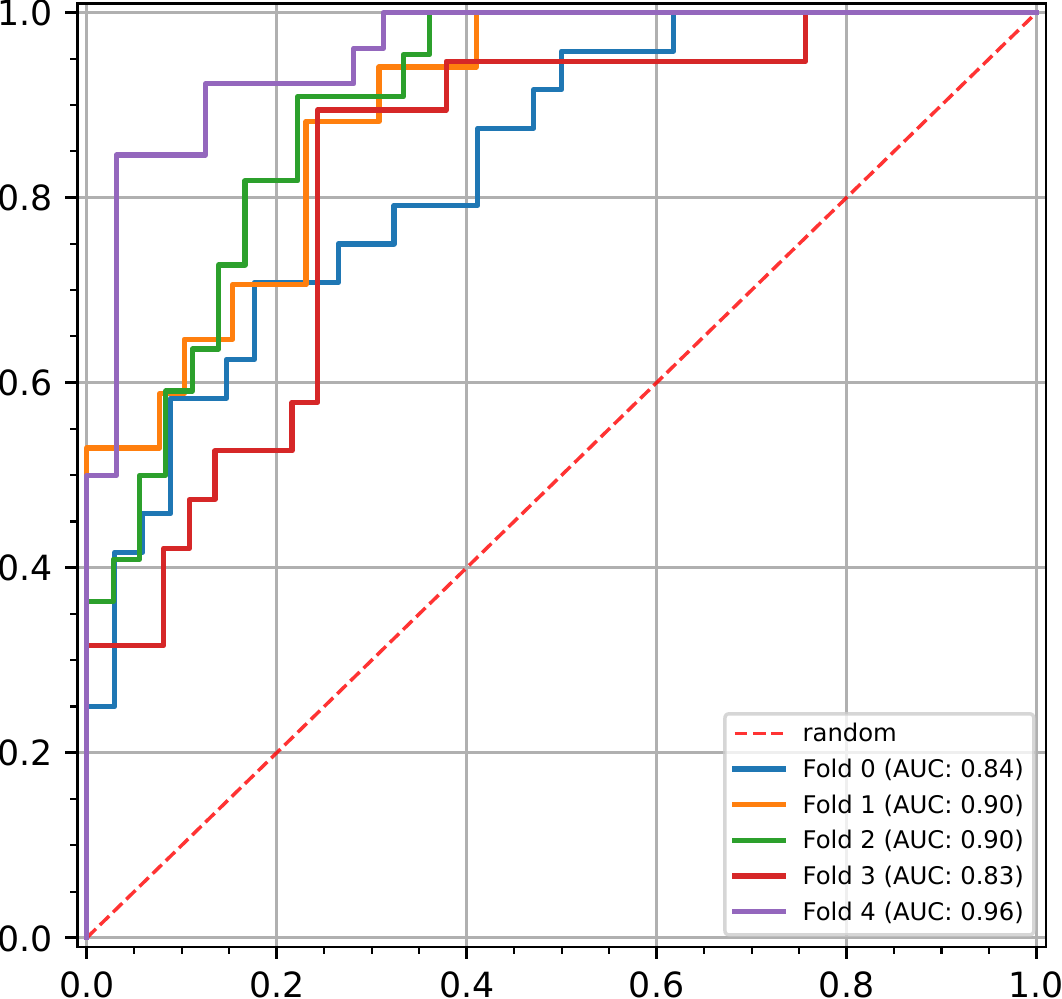}
        \end{subfigure} \vspace{0.3ex} \\
        & \subfloat[]{False positive rate} & & \subfloat[]{False positive rate}
    \end{tabular}
    \caption{\label{fig:results-roc} 
    Receiver Operating Characteristic curves for the 5 folds of (a) IN-RN18 and (b) HistoSSL-RN18 TILMIL model when comparing the predicted sTIL\% score to the sTIL\% score binarized at a cut-off of $20\%$.}
\end{figure}

    


%% file: heatmap_figure.tex
\begin{figure} [ht]
    \centering
    \includegraphics[width=\textwidth]{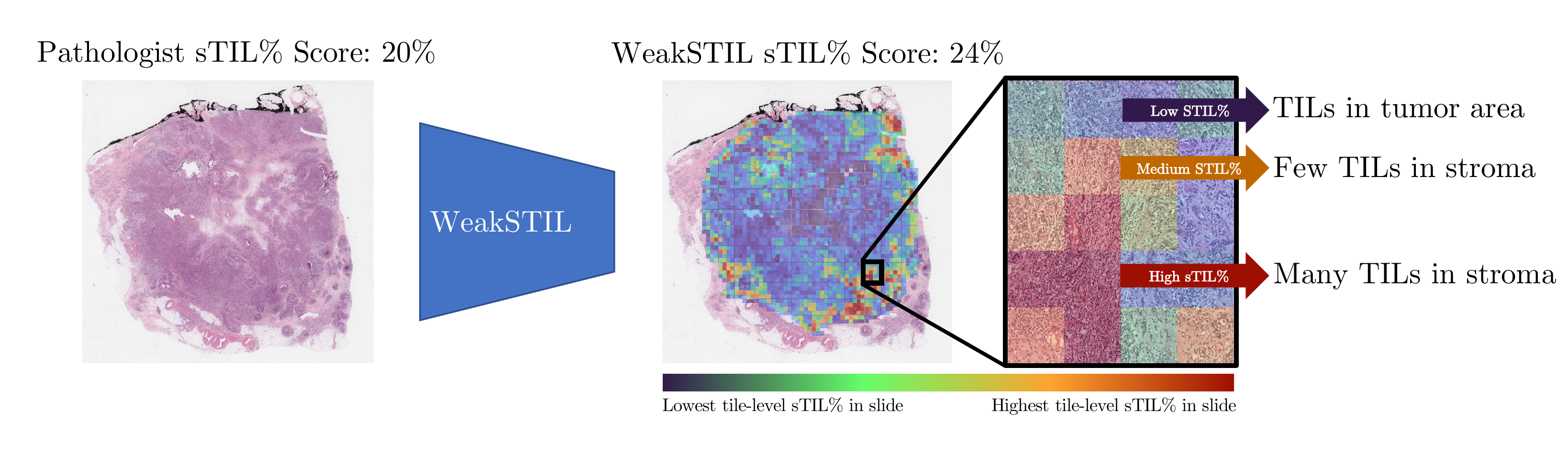}

    \caption{\label{fig:results-visualization} 
    Heatmap visualization of the tile-level sTIL\% scores as predicted by WeakSTIL (HistoSSL-RN18 TILMIL). Left: WeakSTIL receives the tiles of the tumor bed of an H\&E WSI as input, computes an sTIL\% score for each tile, and outputs the mean sTIL\% score of these tiles as WSI-level label score. Right: Close-up of a section of the H\&E WSI. One can see that the model only predicts high sTIL\% scores for TILs that are in stroma, not for TILs that are near tumor cells.
    }
\end{figure}

%% file: hyperparameter_tables.tex
\begin{table*}[ht]
\caption{\label{tab:appendix-table-lr-reg-ssl-linear} Results of hyperparameter grid search using HistoSSL-RN18 TILMIL with a single linear layer ($\boldsymbol{w} \in \mathbb{R}^{512}, b \in \mathbb{R}$) for the sTIL\% regression task. Trained for 50 epochs with a batch size of 1, evaluating every epoch, with varying learning rate and regularization, trained on a subsample of 500 tiles. These results are the 5-fold mean and standard deviation on the validation set.}
    \centering
    \begin{tabular}{crrrrrr}
    \toprule
         \multirow{2}{*}{\textbf{Learning rate}}    & \multicolumn{6}{c}{\textbf{Regularization}} \\ 
                            &      $5\times 10^{-3}$ &         $1\times10^{-3}$&   $5\times10^{-4}$ &         $1\times10^{-4}$ &  $5\times10^{-5}$ & $1\times10^{-5}$ \\
    \midrule
         
            $5\times10^{-2}$&   $72.6\pm1.8$ &   $72.1\pm2.4$  & $75.1\pm4.1$& $82.1\pm4.1$  & $83.1\pm5.8$    &   $81.7\pm6.1$\\
            $1\times10^{-2}$       &   $83.8\pm2.9$ &   $85.5\pm3.6$  & $85.2\pm3.4$& $85.2\pm2.4$  & $85.1\pm2.5$    &   $85.5\pm2.2$\\
            $5\times10^{-3}$&   $82.7\pm3.2$ &   $84.0\pm2.6$  & $84.3\pm2.3$& $85.0\pm1.5$  & $84.7\pm1.5$    &   $84.9\pm1.4$\\
            $1\times10^{-3}$       &   $80.5\pm3.4$ &   $80.7\pm3.2$ &  $80.8\pm3.0$& $81.1\pm2.9$  & $81.3\pm2.9$    &   $81.2\pm2.9$\\
            $5\times10^{-4}$&   $79.5\pm3.6$ &   $79.9\pm3.5$ &  $79.9\pm3.5$& $80.1\pm3.4$  & $80.0\pm3.4$    &   $80.0\pm3.4$\\
            $1\times10^{-4}$       &   $76.9\pm4.1$ &   $77.3\pm4.4$ &  $77.3\pm4.5$& $77.3\pm4.6$  & $77.3\pm4.6$    &   $77.3\pm4.6$\\
            $5\times10^{-5}$&   $74.2\pm5.3$ &   $74.5\pm5.4$ &  $74.5\pm5.4$& $74.5\pm5.4$  & $74.5\pm5.4$    &   $74.5\pm5.4$\\
            $1\times10^{-5}$       &   $64.2\pm8.0$ &   $64.0\pm8.0$ &  $64.0\pm8.0$& $63.9\pm8.0$  & $63.9\pm8.0$    &   $63.9\pm8.0$\\ 
    \bottomrule
    \end{tabular}
\end{table*}

\begin{table*}[ht]
\caption{\label{tab:appendix-table-lr-reg-ssl-non-linear} Results of hyperparameter grid search using HistoSSL-RN18 TILMIL with two linear layers ($\boldsymbol{w}_1 \in \mathbb{R}^{512\times128}, b_1 \in \mathbb{R}^{128}$, $\boldsymbol{w}_2 \in \mathbb{R}^{128}, b_2 \in \mathbb{R}$) for the sTIL\% regression task. Trained for 50 epochs with a batch size of 1, evaluating every epoch, with varying learning rate and regularization, trained on a subsample of 500 tiles. These results are the 5-fold mean and standard deviation on the validation set.}
    \centering
    \begin{tabular}{crrrrrr}
    \toprule
         \multirow{2}{*}{\textbf{Learning rate}}    & \multicolumn{6}{c}{\textbf{Regularization}} \\ 
                            &      $5\times 10^{-3}$ &         $1\times10^{-3}$&   $5\times10^{-4}$ &         $1\times10^{-4}$ &  $5\times10^{-5}$ & $1\times10^{-5}$ \\ 
    \midrule
            $5\times10^{-2}$&   $70.4\pm4.1$ &   $67.4\pm3.3$  & $68.2\pm4.1$& $67.0\pm4.1$  & $69.3\pm6.0$    &   $72.3\pm7.8$\\
            $1\times10^{-2}$       &   $79.5\pm2.4$ &   $76.3\pm6.8$  & $71.5\pm7.1$& $79.2\pm5.6$  & $80.9\pm7.4$    &   $69.2\pm3.1$\\
            $5\times10^{-3}$&   $80.4\pm3.1$ &   $84.8\pm3.0$  & $85.6\pm3.5$& $84.0\pm3.3$  & $84.2\pm3.0$    &   $84.0\pm3.4$\\
            $1\times10^{-3}$       &   $80.2\pm3.6$ &   $83.4\pm3.3$ &  $84.0\pm2.9$& $84.3\pm2.2$  & $84.1\pm2.3$    &   $84.2\pm2.3$\\
            $5\times10^{-4}$&   $80.3\pm3.6$ &   $83.0\pm3.5$ &  $83.3\pm3.2$& $83.5\pm2.7$  & $83.5\pm2.6$    &   $83.3\pm2.6$\\
            $1\times10^{-4}$       &   $79.9\pm3.5$ &   $81.1\pm3.1$ &  $81.3\pm3.0$& $81.7\pm2.7$  & $81.8\pm2.7$    &   $81.7\pm2.6$\\
            $5\times10^{-5}$&   $78.9\pm3.6$ &   $79.7\pm3.3$ &  $80.0\pm3.2$& $80.2\pm3.0$  & $80.2\pm3.0$    &   $80.2\pm3.1$\\
            $1\times10^{-5}$       &   $75.1\pm4.9$ &   $75.9\pm4.7$ &  $75.9\pm4.7$& $76.0\pm4.7$  & $76.0\pm4.7$    &   $76.0\pm4.6$\\ 
    \bottomrule    
    \end{tabular}
\end{table*}

\begin{table*}[ht]
\caption{\label{tab:appendix-table-lr-reg-ssl-double-head} Results of hyperparameter grid search using HistoSSL-RN18 TILMIL with two linear layers and a tanh non-linearity after the first layer ($\boldsymbol{w}_1 \in \mathbb{R}^{512\times128}, b_1 \in \mathbb{R}^{128}$, $\boldsymbol{w}_2 \in \mathbb{R}^{128}, b_2 \in \mathbb{R}$) for the sTIL\% regression task. Trained for 50 epochs with a batch size of 1, evaluating every epoch, with varying learning rate and regularization, trained on a subsample of 500 tiles. These results are the 5-fold mean and standard deviation on the validation set.}
    \centering
    \begin{tabular}{crrrrrr}
    \toprule
         \multirow{2}{*}{\textbf{Learning rate}}    & \multicolumn{6}{c}{\textbf{Regularization}} \\ 
                            &      $5\times 10^{-3}$ &         $1\times10^{-3}$&   $5\times10^{-4}$ &         $1\times10^{-4}$ &  $5\times10^{-5}$ & $1\times10^{-5}$ \\
    \midrule
            $5\times10^{-2}$&   $70.8\pm2.7$ &   $69.0\pm2.6$  & $70.9\pm2.0$& $68.5\pm1.4$  & $68.5\pm3.3$    &   $69.0\pm5.6$\\
            $1\times10^{-2}$       &   $73.5\pm1.9$ &   $76.8\pm2.3$  & $76.2\pm2.5$& $84.6\pm4.3$  & $84.6\pm3.2$    &   $83.3\pm3.1$\\
            $5\times10^{-3}$&   $75.0\pm3.6$ &   $85.2\pm4.4$  & $86.3\pm3.8$& $84.7\pm3.6$  & $86.0\pm3.4$    &   $85.6\pm4.0$\\
            $1\times10^{-3}$       &   $80.6\pm3.0$ &   $83.9\pm3.0$ &  $84.4\pm3.2$& $84.3\pm2.9$  & $84.3\pm2.9$    &   $84.4\pm2.6$\\
            $5\times10^{-4}$&   $80.5\pm3.3$ &   $83.3\pm3.6$ &  $83.5\pm3.4$& $83.5\pm2.8$  & $83.5\pm2.6$    &   $83.8\pm2.5$\\
            $1\times10^{-4}$       &   $79.9\pm3.5$ &   $81.1\pm3.1$ &  $81.3\pm3.0$& $81.6\pm2.7$  & $81.6\pm2.7$    &   $81.8\pm2.7$\\
            $5\times10^{-5}$&   $78.9\pm3.6$ &   $79.7\pm3.4$ &  $79.9\pm3.3$& $80.2\pm3.2$  & $80.2\pm3.2$    &   $80.3\pm3.1$\\
            $1\times10^{-5}$       &   $75.1\pm4.9$ &   $75.9\pm4.8$ &  $76.0\pm4.8$& $76.1\pm4.8$  & $76.1\pm4.8$    &   $76.1\pm4.8$\\ 
    \bottomrule       
    \end{tabular}
\end{table*}

\begin{table*}[ht]
\caption{\label{tab:appendix-table-lr-reg-imagenet-linear} Results of hyperparameter grid search using IN-RN18 TILMIL with a single linear layer ($\boldsymbol{w} \in \mathbb{R}^{512}, b \in \mathbb{R}$) for the sTIL\% regression task. Trained for 50 epochs with a batch size of 1, evaluating every epoch, with varying learning rate and regularization, trained on a subsample of 500 tiles. These results are the 5-fold mean and standard deviation on the validation set.}
    \centering
    \begin{tabular}{crrrrrr}
    \toprule
         \multirow{2}{*}{\textbf{Learning rate}}    & \multicolumn{6}{c}{\textbf{Regularization}} \\ 
                            &      $5\times 10^{-3}$ &         $1\times10^{-3}$&   $5\times10^{-4}$ &         $1\times10^{-4}$ &  $5\times10^{-5}$ & $1\times10^{-5}$ \\ 
    \midrule
            $5\times10^{-2}$&   $65.5\pm5.5$ &   $66.1\pm5.0$  & $65.3\pm4.8$& $65.1\pm5.0$  & $62.9\pm5.2$    &   $65.2\pm5.8$\\
            $1\times10^{-2}$       &   $65.7\pm5.1$ &   $65.4\pm5.2$  & $65.5\pm4.7$& $64.2\pm5.6$  & $64.3\pm5.1$    &   $63.7\pm4.9$\\
            $5\times10^{-3}$&   $65.9\pm5.6$ &   $64.5\pm5.5$  & $63.3\pm6.1$& $62.7\pm5.4$  & $62.4\pm5.6$    &   $61.7\pm6.5$\\
            $1\times10^{-3}$       &   $58.0\pm6.8$ &   $59.1\pm6.9$ &  $59.2\pm6.9$& $59.3\pm6.9$  & $59.3\pm6.9$    &   $59.3\pm6.9$\\
            $5\times10^{-4}$&   $53.3\pm5.1$ &   $53.2\pm4.9$ &  $53.2\pm4.8$& $53.2\pm4.9$  & $53.2\pm4.9$    &   $53.1\pm4.9$\\
            $1\times10^{-4}$       &   $51.9\pm4.7$ &   $51.5\pm4.6$ &  $51.4\pm4.6$& $51.4\pm4.6$  & $51.4\pm4.6$    &   $51.4\pm4.6$\\
            $5\times10^{-5}$&   $50.4\pm4.3$ &   $50.4\pm4.4$ &  $50.3\pm4.3$& $50.3\pm4.3$  & $50.3\pm4.3$    &   $50.3\pm4.3$\\
            $1\times10^{-5}$       &   $49.7\pm3.1$ &   $49.7\pm3.1$ &  $49.7\pm3.1$& $49.7\pm3.1$  & $49.7\pm3.1$    &   $49.7\pm3.1$\\ 
    \bottomrule
    \end{tabular}
\end{table*}

%% file: fig-heatmap-appendix-success-1.tex
\begin{figure} [ht]
    \centering
    \includegraphics[width=\textwidth]{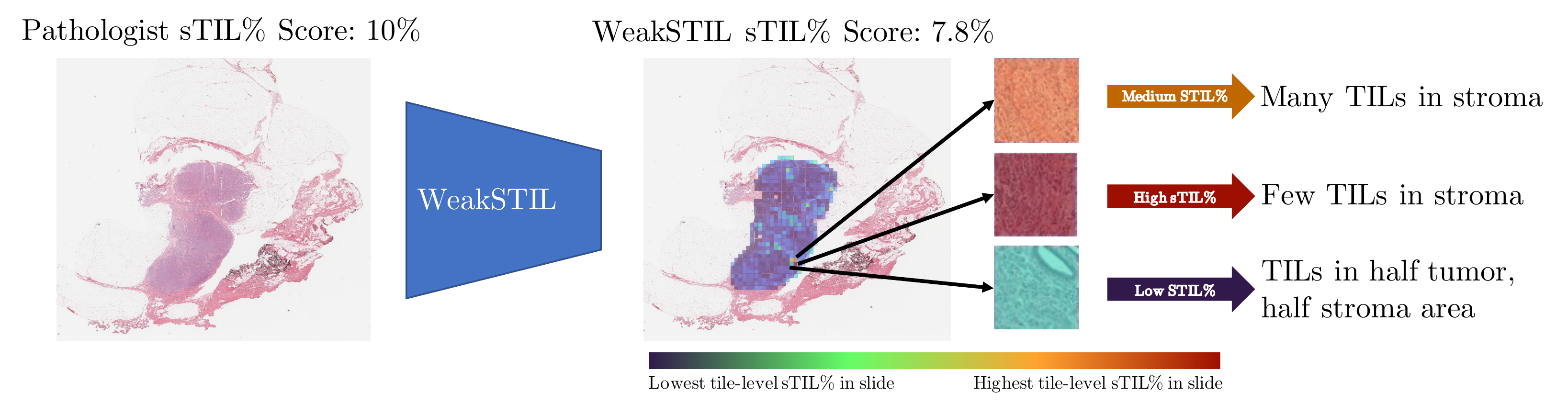}
    \caption{\label{fig:apx-viz-success-1} 
    Heatmap visualization of the tile-level sTIL\% scores as predicted by WeakSTIL (HistoSSL-RN18 TILMIL). Left: WeakSTIL receives the tiles of the tumor bed of an H\&E WSI as input, computes an sTIL\% score for each tile, and outputs the mean sTIL\% score of these tiles as WSI-level label score. Right: Close-up of a section of the H\&E WSI. We see a predicted sTIL\% score that is close to the pathologist's score with sensible tile-level predictions.
    }
\end{figure}

%% file: fig-heatmap-appendix-success-2.tex
\begin{figure} [!ht]
    \centering
    \includegraphics[width=\textwidth]{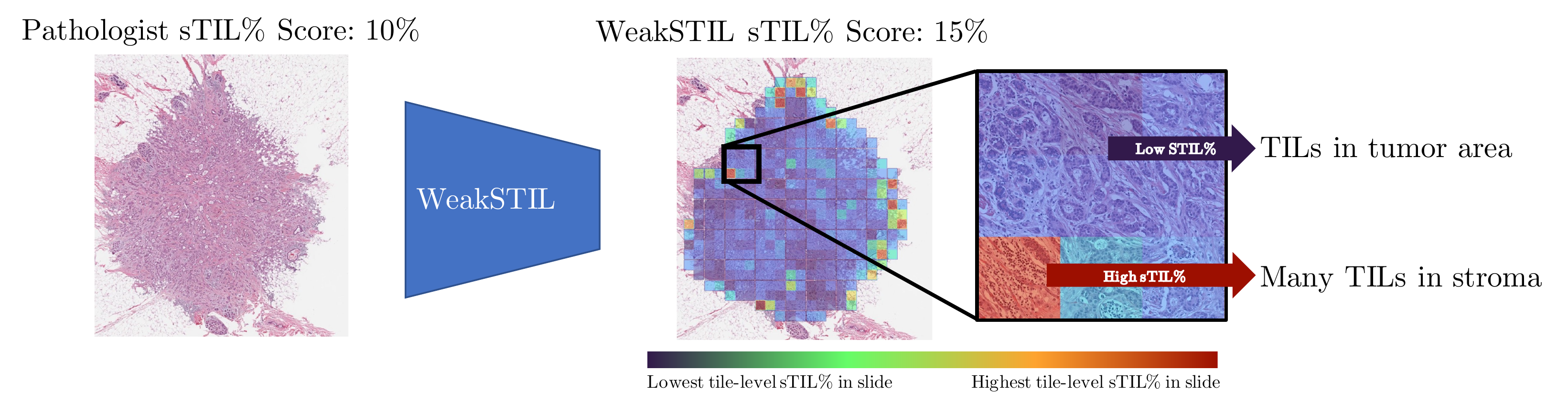}
    \caption{\label{fig:apx-viz-success-2} 
    Heatmap visualization of the tile-level sTIL\% scores as predicted by WeakSTIL (HistoSSL-RN18 TILMIL). Left: WeakSTIL receives the tiles of the tumor bed of an H\&E WSI as input, computes an sTIL\% score for each tile, and outputs the mean sTIL\% score of these tiles as WSI-level label score. Right: Close-up of a section of the H\&E WSI. We see a predicted sTIL\% score that is close to the pathologist's score with sensible tile-level predictions.
    }
\end{figure}

%% file: fig-heatmap-appendix-fail-high-stil.tex
\begin{figure} [!ht]
    \centering
    \includegraphics[width=\textwidth]{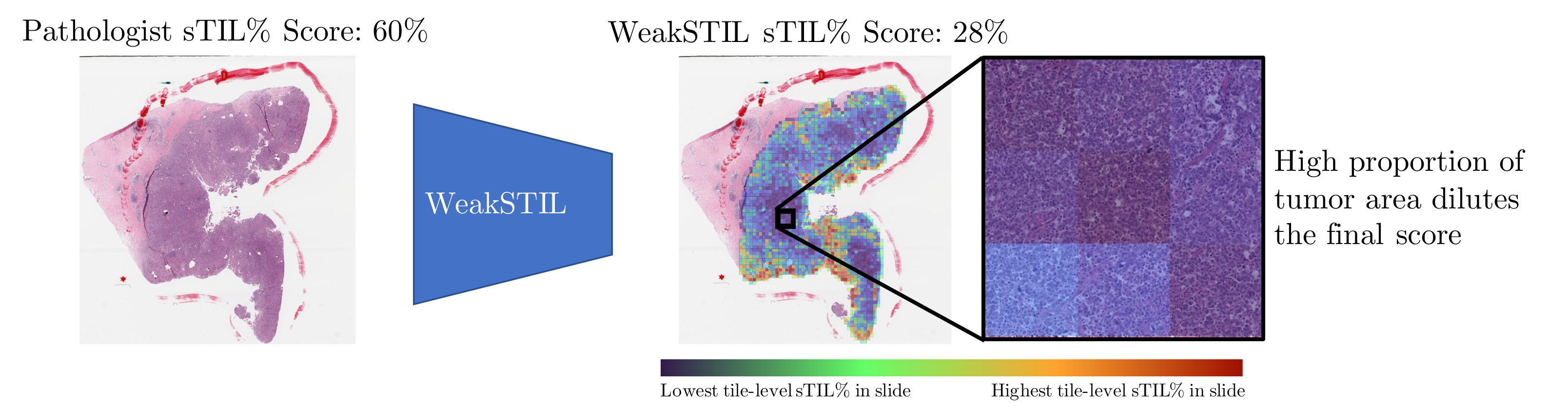}
    \caption{\label{fig:apx-viz-fail-high} 
    Heatmap visualization of the tile-level sTIL\% scores as predicted by WeakSTIL (HistoSSL-RN18 TILMIL). Left: WeakSTIL receives the tiles of the tumor bed of an H\&E WSI as input, computes an sTIL\% score for each tile, and outputs the mean sTIL\% score of these tiles as WSI-level label score. Right: Close-up of a section of the H\&E WSI. One can see that the tumor bed of the tissue sample contains mostly tumor area. The model correctly predicts sTIL\% scores for these tumor tiles, but since the mean is computed over all tumor bed tiles, the final WSI-level score is too low.
    }
\end{figure}

%% file: fig-heatmap-appendix-fail-edge.tex
\begin{figure} [!ht]
    \centering
    \includegraphics[width=\textwidth]{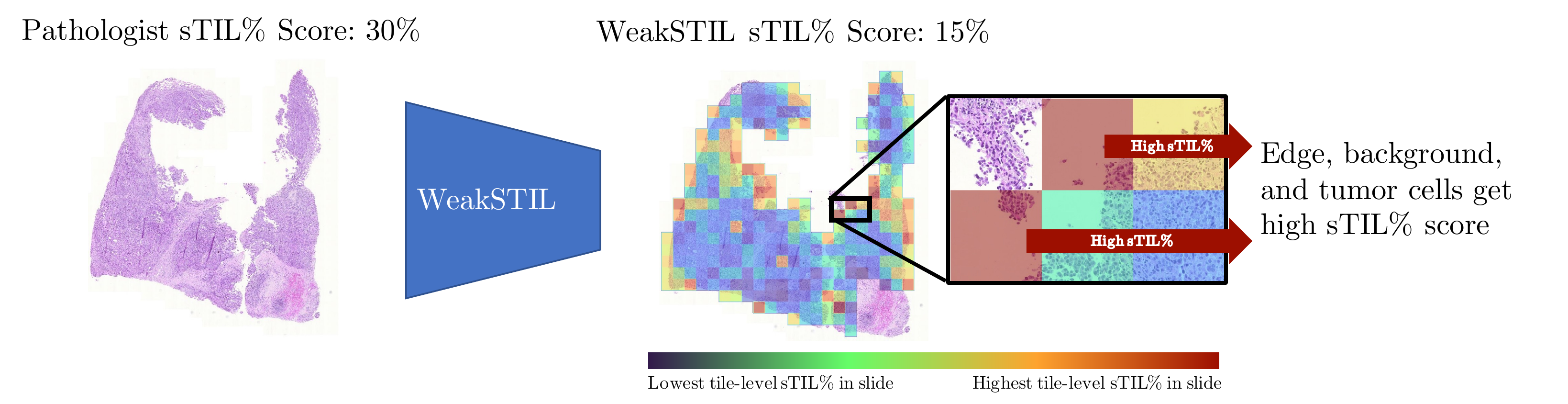}
    \caption{\label{fig:apx-viz-fail-edge} 
   Heatmap visualization of the tile-level sTIL\% scores as predicted by WeakSTIL (HistoSSL-RN18 TILMIL). Left: WeakSTIL receives the tiles of the tumor bed of an H\&E WSI as input, computes an sTIL\% score for each tile, and outputs the mean sTIL\% score of these tiles as WSI-level label score. Right: Close-up of a section of the H\&E WSI. One can see that the model predicts high sTIL\% scores for edge tiles that do not contain stroma or TILs.
    }
\end{figure}